\documentclass{elsart}
\usepackage{epsfig}
\usepackage{amssymb}
\usepackage{amsmath}

\newcommand{\affuni}[2]{Dipartimento di Fisica dell'Universit\`a #1, #2, Italy.}
\newcommand{\affinfn}[2]{INFN Sezione di #1, #2, Italy.}

\newcommand{\dafne}     {DA$\Phi$NE }

\newcommand{\ep}{\mbox{$e^{+}$}}
\newcommand{\el}{\mbox{$e^{-}$}}
\newcommand{\pio}{\mbox{$\pi^{0}$}}
\newcommand{\pim}{\mbox{$\pi^{-}$}}
\newcommand{\pip}{\mbox{$\pi^{+}$}}
\newcommand{\ks}{\mbox{$K_{S}$}}
\newcommand{\kl}{\mbox{$K_{L}$}}

\begin{document}

\begin{frontmatter}

\title{\boldmath Limit on the production of a light vector gauge boson in 
  $\phi$ meson decays with the KLOE detector}

\collab{The KLOE-2 Collaboration}
\author[Frascati]{D.~Babusci},
\author[Roma2,INFNRoma2]{D.~Badoni},
\author[Cracow]{I.~Balwierz-Pytko},
\author[Frascati]{G.~Bencivenni},
\author[Roma1,INFNRoma1]{C.~Bini},
\author[Frascati]{C.~Bloise},
\author[Frascati]{F.~Bossi},
\author[INFNRoma3]{P.~Branchini},
\author[Roma3,INFNRoma3]{A.~Budano},
\author[Uppsala]{L.~Caldeira~Balkest\aa hl},
\author[Frascati]{G.~Capon},
\author[Roma3,INFNRoma3]{F.~Ceradini},
\author[Frascati]{P.~Ciambrone},
\author[Cracow]{E.~Czerwi\'nski},
\author[Frascati]{E.~Dan\`e},
\author[Frascati]{E.~De~Lucia},
\author[INFNBari]{G.~De~Robertis},
\author[Roma1,INFNRoma1]{A.~De~Santis},
\author[Roma1,INFNRoma1]{A.~Di~Domenico},
\author[Napoli,INFNNapoli]{C.~Di~Donato},
\author[INFNRoma2]{R.~Di~Salvo},
\author[Frascati]{D.~Domenici},
\author[Bari,INFNBari]{O.~Erriquez},
\author[Bari,INFNBari]{G.~Fanizzi},
\author[Roma2,INFNRoma2]{A.~Fantini},
\author[Frascati]{G.~Felici},
\author[Roma1,INFNRoma1]{S.~Fiore},
\author[Roma1,INFNRoma1]{P.~Franzini},
\author[Roma1,INFNRoma1]{P.~Gauzzi},
\author[Messina,INFNCatania]{G.~Giardina},
\author[Frascati]{S.~Giovannella\corauthref{cor}},
\ead{simona.giovannella@lnf.infn.it}
\corauth[cor]{Corresponding author.}
\author[Roma2,INFNRoma2]{F.~Gonnella},
\author[INFNRoma3]{E.~Graziani},
\author[Frascati]{F.~Happacher},
\author[Uppsala]{L.~Heijkenskj\"old}
\author[Uppsala]{B.~H\"oistad},
\author[Frascati]{L.~Iafolla},
\author[Uppsala]{M.~Jacewicz},
\author[Uppsala]{T.~Johansson},
\author[Uppsala]{A.~Kupsc},
\author[Frascati,StonyBrook]{J.~Lee-Franzini},
\author[Frascati]{B.~Leverington},
\author[INFNBari]{F.~Loddo},
\author[Roma3,INFNRoma3]{S.~Loffredo},
\author[Messina,INFNCatania,CentroCatania]{G.~Mandaglio},
\author[Moscow]{M.~Martemianov},
\author[Frascati,Marconi]{M.~Martini},
\author[Roma2,INFNRoma2]{M.~Mascolo},
\author[Roma2,INFNRoma2]{R.~Messi},
\author[Frascati]{S.~Miscetti},
\author[Frascati]{G.~Morello},
\author[INFNRoma2]{D.~Moricciani},
\author[Cracow]{P.~Moskal},
\author[INFNRoma3,LIP]{F.~Nguyen},
\author[INFNRoma3]{A.~Passeri},
\author[Energetica,Frascati]{V.~Patera},
\author[Roma3,INFNRoma3]{I.~Prado~Longhi},
\author[INFNBari]{A.~Ranieri},
\author[Mainz]{C.~F.~Redmer},
\author[Frascati]{P.~Santangelo},
\author[Frascati]{I.~Sarra\corauthref{cor}},
\ead{ivano.sarra@lnf.infn.it}
\author[Calabria,INFNCalabria]{M.~Schioppa},
\author[Frascati]{B.~Sciascia},
\author[Cracow]{M.~Silarski},
\author[Roma3,INFNRoma3]{C.~Taccini},
\author[INFNRoma3]{L.~Tortora},
\author[Frascati]{G.~Venanzoni},
\author[Warsaw]{W.~Wi\'slicki},
\author[Uppsala]{M.~Wolke},
\author[Cracow]{J.~Zdebik}

\address[Bari]{\affuni{di Bari}{Bari}}
\address[INFNBari]{\affinfn{Bari}{Bari}}
\address[CentroCatania]{Centro Siciliano di Fisica Nucleare e Struttura della Materia, Catania, Italy.}
\address[INFNCatania]{\affinfn{Catania}{Catania}}
\address[Calabria]{\affuni{della Calabria}{Cosenza}}
\address[INFNCalabria]{INFN Gruppo collegato di Cosenza, Cosenza, Italy.}
\address[Cracow]{Institute of Physics, Jagiellonian University, Cracow, Poland.}
\address[Frascati]{Laboratori Nazionali di Frascati dell'INFN, Frascati, Italy.}
\address[Mainz]{Institut f\"ur Kernphysik, 
Johannes Gutenberg Universit\"at Mainz, Germany.}
\address[Messina]{Dipartimento di Fisica e Scienze della Terra dell'Universit\`a di Messina, Messina, Italy.}\address[Moscow]{Institute for Theoretical and Experimental Physics (ITEP), Moscow, Russia.}
\address[Napoli]{\affuni{``Federico II''}{Napoli}}
\address[INFNNapoli]{\affinfn{Napoli}{Napoli}}
\address[Energetica]{Dipartimento di Scienze di Base ed Applicate per l'Ingegneria dell'Universit\`a 
``Sapienza'', Roma, Italy.}
\address[Marconi]{Dipartimento di Scienze e Tecnologie applicate, Universit\`a ``Guglielmo Marconi", Roma, Italy.}
\address[Roma1]{\affuni{``Sapienza''}{Roma}}
\address[INFNRoma1]{\affinfn{Roma}{Roma}}
\address[Roma2]{\affuni{``Tor Vergata''}{Roma}}
\address[INFNRoma2]{\affinfn{Roma Tor Vergata}{Roma}}
\address[Roma3]{\affuni{``Roma Tre''}{Roma}}
\address[INFNRoma3]{\affinfn{Roma Tre}{Roma}}
\address[StonyBrook]{Physics Department, State University of New 
York at Stony Brook, USA.}
\address[Uppsala]{Department of Physics and Astronomy, Uppsala University, Uppsala, Sweden.}
\address[Warsaw]{National Centre for Nuclear Research, Warsaw, Poland.}
\address[LIP]{Present Address: Laborat\'orio de Instrumenta\c{c}\~{a}o e F\'isica Experimental de Part\'iculas,
Lisbon, Portugal.}


\begin{abstract}

We present a new limit on the production of a light dark-force
mediator with the KLOE detector at DA$\Phi$NE. This boson,
called $U$, has been searched for in the decay $\phi\to\eta\,U$,
$U\to\ep\el$, analyzing the decay $\eta\to\pio\pio\pio$ in a data 
sample of 1.7 fb$^{-1}$.
No structures are observed in the \ep\el\ invariant mass distribution 
over the background.
This search is combined with a previous result obtained from the 
decay $\eta\to\pip\pim\pio$, increasing the sensitivity. We set
an upper limit at 90\% C.L.\  
on the ratio between the $U$ boson coupling constant and the fine 
structure constant of 
$\alpha'/\alpha < 1.7 \times 10^{-5}$ for $30<M_{U}<400$ MeV and
$\alpha'/\alpha \leq 8 \times 10^{-6}$ for the sub-region 
$50<M_{U}<210$ MeV. 
This result assumes the Vector Meson Dominance expectations for the 
$\phi\eta\gamma^*$ transition form factor. The dependence of this limit 
on the transition form factor has also been studied.

\end{abstract}


\begin{keyword}
$e^{+}e^{-}$ collisions \sep dark forces \sep gauge vector boson

\PACS 14.70.Pw  
\end{keyword}

\end{frontmatter}

\section{Introduction}
\label{Sec:Intro}

There are theories beyond the Standard Model (SM) that postulate the 
existence of new bosons, mediators of some hidden gauge group, under 
which ordinary matter is uncharged \cite{Fayet,Batell:2009di}.
Such theories have been recently invoked to account for some puzzling 
astrophysical observations and predict
that at least one of the new bosons is neutral, 
relatively light and weakly coupled with ordinary matter 
\cite{ArkaniHamed:2008qp,Pospelov:2007mp}. The coupling is 
represented by adding to the Lagrangian a kinetic mixing 
term \cite{Holdom} between 
the new ($U$) boson and the photon, whose strength is 
measured by a parameter $\epsilon$ that could be as large as 
$\sim 10^{-3}$. Due to this mixing, the $U$ boson can be produced in 
processes involving SM particles, and can decay into \ep\el, 
$\mu^+\mu^-$, $\pi^+\pi^-$..., depending on its mass and 
on model-specific details.

Searches for the $U$ boson have been recently performed at $e$-$p$ 
fixed target facilities \cite{MAMI,APEX}, and, in conjunction with 
a dark Higgs, at \ep\el\ colliders \cite{BaBar}, with null results.
The $U$ can also be searched for in vector ($V$) to pseudoscalar ($P$) 
meson decays, with a rate that is $\epsilon^2$ times suppressed with 
respect to the ordinary $V \rightarrow P \gamma$ transitions 
\cite{Reece:2009un}. Since the $U$ is supposed to decay to \ep\el\ 
with a non-negligible branching ratio, $V \rightarrow P U$ events 
will produce a sharp peak in the invariant mass distribution of the 
electron-positron pair over the continuum background due to Dalitz 
decay events $V \to P \ep\el$. Using this approach, KLOE has already 
published a limit on the existence of the $U$ boson \cite{UbosonKLOE}, 
studying $\phi\to\eta\,\ep\el$ decays, where the $\eta$ meson was 
tagged by its $\pip\pim\pio$ decay.
In this letter, we present an update of this analysis, which improves 
background rejection for the already used data sample (1.5 fb$^{-1}$) 
and increases statistics by a factor of about three, exploiting also 
the neutral $\eta\to\pio\pio\pio$ decay chain, with a data sample of 
1.7 fb $^{-1}$.

\section{The KLOE detector}

\dafne, the Frascati $\phi$-factory, is an $e^+e^-$ collider running at 
center-of-mass energy of $\sim 1020$~MeV. 
Positron and electron beams collide at an angle of $\pi$-25 mrad, 
producing $\phi$ mesons nearly at rest. The KLOE experiment operated at 
this collider from 2000 to 2006, collecting 2.5 fb$^{-1}$.
The KLOE detector consists of a large cylindrical Drift Chamber (DC),
surrounded by a lead-scintillating fiber electromagnetic calorimeter
(EMC), all embedded inside a superconducting coil, providing a 0.52~T 
axial field.
The beam pipe at the interaction region is a sphere with 10 cm radius, 
made of a 0.5 mm thick Beryllium-Aluminum alloy. 
The drift chamber~\cite{DCH}, 4~m in diameter and 3.3~m long, has 
12,582 all-stereo tungsten sense wires and 37,746 aluminum field wires,
with a shell made of carbon fiber-epoxy composite with an internal wall 
of $\sim 1$ mm thickness. The gas used is a 90\% helium, 10\% isobutane 
mixture. The momentum resolution is $\sigma(p_{\perp})/p_{\perp}\approx 0.4\%$.
Vertices are reconstructed with a spatial resolution of $\sim$ 3~mm.
The calorimeter~\cite{EMC}, with a readout granularity of 
$\sim$\,(4.4 $\times$ 4.4)~cm$^2$, for a total of 2440 cells arranged 
in five layers, covers 98\% of the solid angle. Each cell is read out 
at both ends by photomultipliers, both in amplitude and time. 
The energy deposits are obtained from the signal amplitude while the
arrival times and the particles positions are obtained from the time
differences. 
Cells close in time and space are grouped into energy clusters. 
Energy and time resolutions are $\sigma_E/E = 5.7\%/\sqrt{E\ {\rm(GeV)}}$ 
and  $\sigma_t = 57\ {\rm ps}/\sqrt{E\ {\rm(GeV)}} \oplus100\ {\rm ps}$, 
respectively.
The trigger \cite{TRG} uses both calorimeter and chamber information.
In this analysis the events are selected by the calorimeter trigger,
requiring two energy deposits with $E>50$ MeV for the barrel and $E>150$
MeV for the endcaps. 
Data are then analyzed by an event classification filter \cite{NIMOffline},
which selects and streams various categories of events in different
output files.

\section{\boldmath Event selection}
\label{Sec:DataSample}

To improve the search for the $U$ boson, we have carried out the 
analysis of the process $\phi\to\eta\,U$, $U\to e^+ e^-$, adding the 
decay channel $\eta\to\pio\pio\pio$ to the previously used, 
$\eta\to\pip\pim\pio$. 
The new search has been performed on a data sample of 1.7 fb$^{-1}$, 
corresponding approximately to $6\times 10^9$ produced $\phi$ mesons. 
The Monte Carlo (MC) simulation for the $\phi\to\eta\, U$ decay has 
been developed according to \cite{Reece:2009un}, with a flat distribution 
in the \ep\el\ invariant mass, $M_{ee}$, while the irreducible background 
$\phi\to\eta\,\ep\el$, $\eta\to\pi\pi\pi$, has been simulated according 
to a Vector Meson Dominance parametrization \cite{Landsberg85}.
All MC productions, including all other $\phi$ decays, take into account 
changes in \dafne\ operation and background conditions on a run-by-run 
basis. Corrections for data-MC discrepancies in cluster energies and 
tracking efficiency, evaluated with radiative Bhabha scattering 
and $\phi\to\rho\pi$ event samples, respectively, have been applied.

\begin{figure}[!t]
  \begin{center}
    \epsfig{file=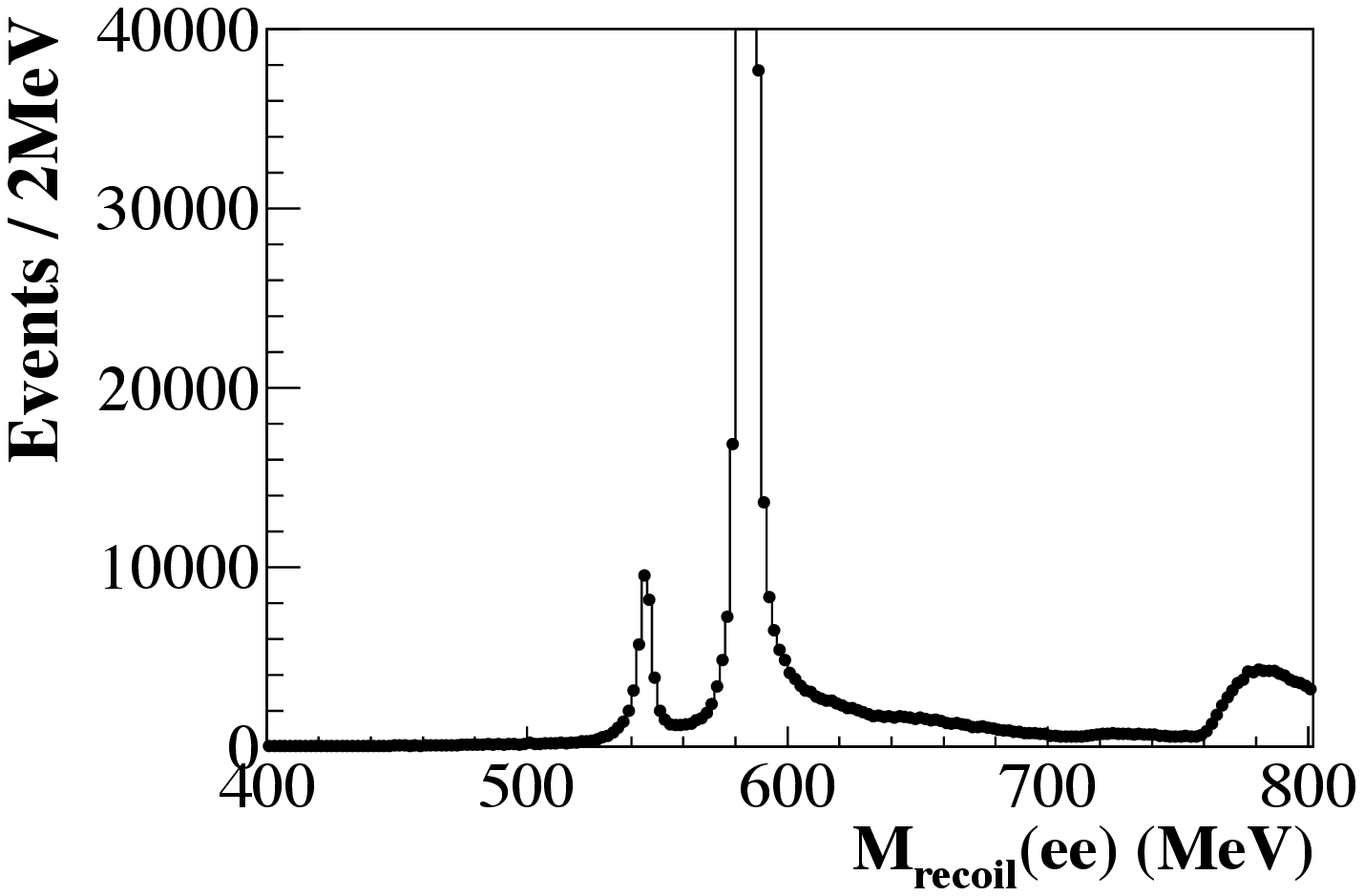,width=0.6\textwidth,angle=90}
  \end{center}
  \caption{Recoiling mass against the $e^+e^-$ pair for the data 
    sample after preselection cuts. The $\phi\to\eta\,e^+e^-$ signal 
    is clearly visible as the peak corresponding to the $\eta$ mass.}
  \label{Fig:Mmiss}
\end{figure}

\begin{figure}[!t]
  \begin{center}
    \epsfig{file=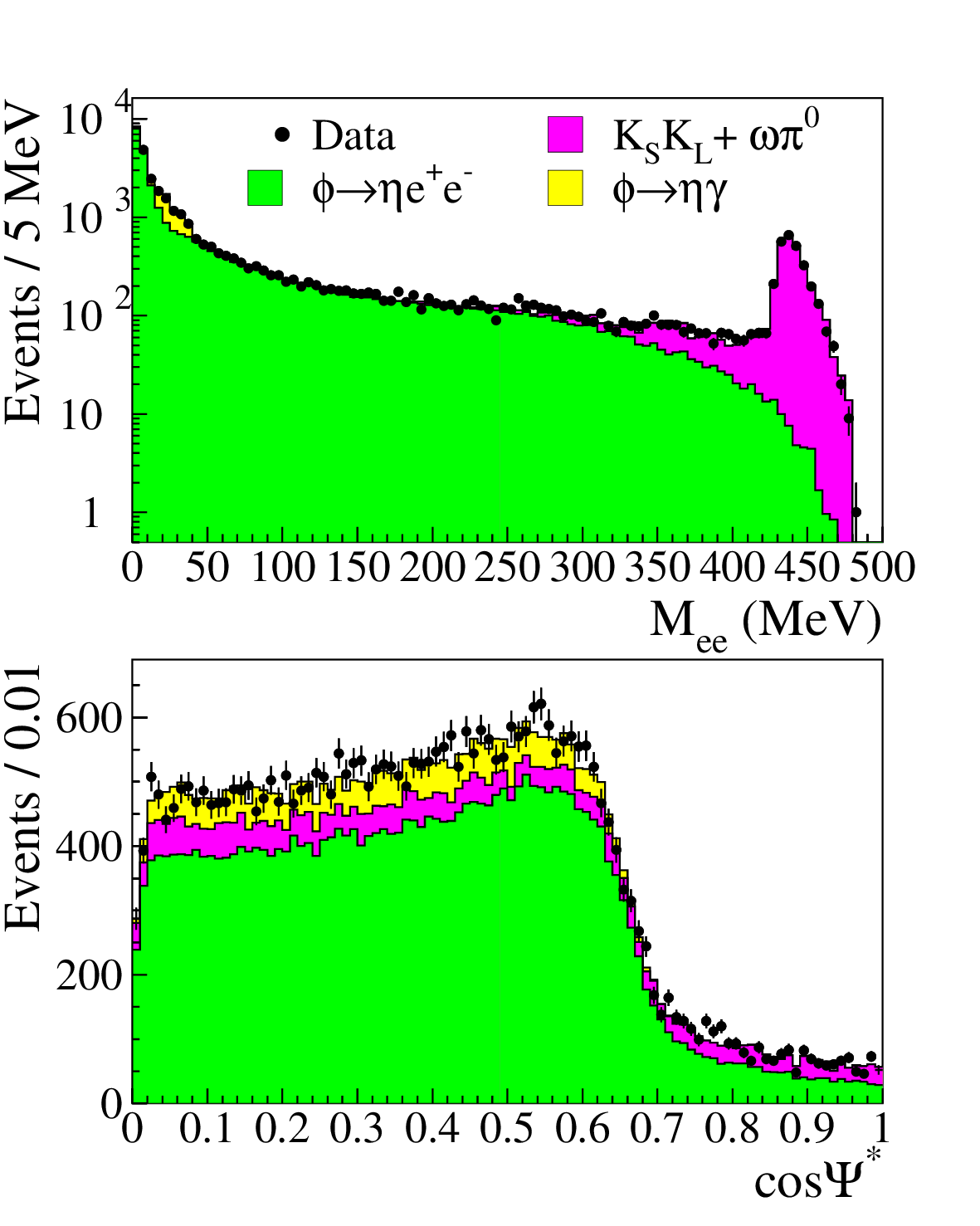,width=0.6\textwidth}
  \end{center}
  \caption{$\phi\to\eta\,\ep\el$, $\eta\to\pio\pio\pio$ events: 
    data-MC comparison for $M_{ee}$ (top) and $\cos\Psi^*$ 
    distributions (bottom) after the $M_{\rm recoil}(ee)$ cut.}
  \label{Fig:CompNeuBefore}
\end{figure}

As a first analysis step for the neutral $\eta$ decay channel, a 
preselection is performed requiring:
\begin{enumerate}
\item two opposite charge tracks with point of closest approach to 
  the beam line inside a cylinder around the interaction point (IP), 
  of 4 cm transverse radius and 20 cm length;
\item six prompt photon candidates, {\it i.e.} energy clusters with
  $E > 7$ MeV not associated to any track, in an angular acceptance 
  $|\cos\theta_\gamma|<0.92$ and in the expected time window for a 
  photon ($|T_\gamma-R_\gamma/c|<{\rm MIN}(3\sigma_t,2\,{\rm ns})$); 
\item a loose cut on the six-photon invariant mass:
  $400 < M_{6\gamma} < 700$ MeV.
\end{enumerate}
After this selection, a peak corresponding to the $\eta$ mass 
is clearly observed in the distribution of the recoil mass against 
the $e^+e^-$ pair, $M_{\rm recoil}(ee)$ (Fig.~\ref{Fig:Mmiss}). The 
second peak at $\sim 590$ MeV is due to $\ks\to\pi^+\pi^-$ decays 
with wrong mass assignment. To select $\phi\to\eta\,\ep\el$ events, 
a $3\sigma$ cut is applied on this variable, 
$536.5 < M_{\rm recoil}(ee) < 554.5$ MeV. The retained sample has 
$\sim 20\%$ residual background contamination, constituted by 
$\phi\to\eta\gamma$, $\phi\to\ks\kl$ and $\ep\el\to\omega\pio$
(about 50\%, 35\% and 15\% of the whole background contribution, 
respectively).
In Fig.~\ref{Fig:CompNeuBefore}, the comparison between data and 
Monte Carlo events for the $M_{ee}$ and $\cos\Psi^*$ distributions 
is shown at this analysis level. The $\Psi^*$ variable is the angle 
between the directions of the $\eta$ and the $e^+$ in the $e^+e^-$ 
rest frame. 
Photon conversion events are concentrated at $M_{ee} \sim 30$ MeV
and $\cos\Psi^*<0.6$, while the other backgrounds cover the
$M_{ee} > 300$ MeV region and are uniformly distributed in $\cos\Psi^*$.

The $\phi\to\eta\gamma$ background contamination is mainly due to
events where a photon converts to an $\ep\el$ pair on the
beam pipe (BP) or drift chamber walls (DCW). After tracing back 
the tracks of the two $e^+$/$e^-$ candidates, these events are 
efficiently rejected by reconstructing the invariant mass ($M_{ee}$) 
and the distance ($D_{ee}$) of the track pair both at the BP and 
DCW surfaces. Both variables are expected to be small for photon 
conversion events, so that this background is removed by 
rejecting events with: 
[~$M_{ee}(BP) <  10$ MeV and $D_{ee}(BP) < 2$ cm~] or
[~$M_{ee}(DCW)< 120$ MeV and $D_{ee}(DCW)< 4$ cm~].

At this stage of the analysis, the surviving background is dominated
by events with two charged pions in the final state, and it is rejected 
by exploiting the timing capabilities of the calorimeter.
When an energy cluster is associated to a track, the time of flight (ToF) 
to the calorimeter is evaluated both using the track trajectory 
($T_{\rm track}=L_{\rm track}/\beta c$) and the calorimeter timing 
($T_{\rm cluster}$). The $\Delta T = T_{\rm track}-T_{\rm cluster}$ variable 
is then evaluated in the electron hypothesis ($\Delta T_e$). In order  
to be fully efficient on signal, events with either an \ep\ or an \el\ 
candidate inside a $3\sigma$ window around $\Delta T_e = 0$
are retained for further analysis.

\begin{figure}[!t]
  \begin{center}
    \epsfig{file=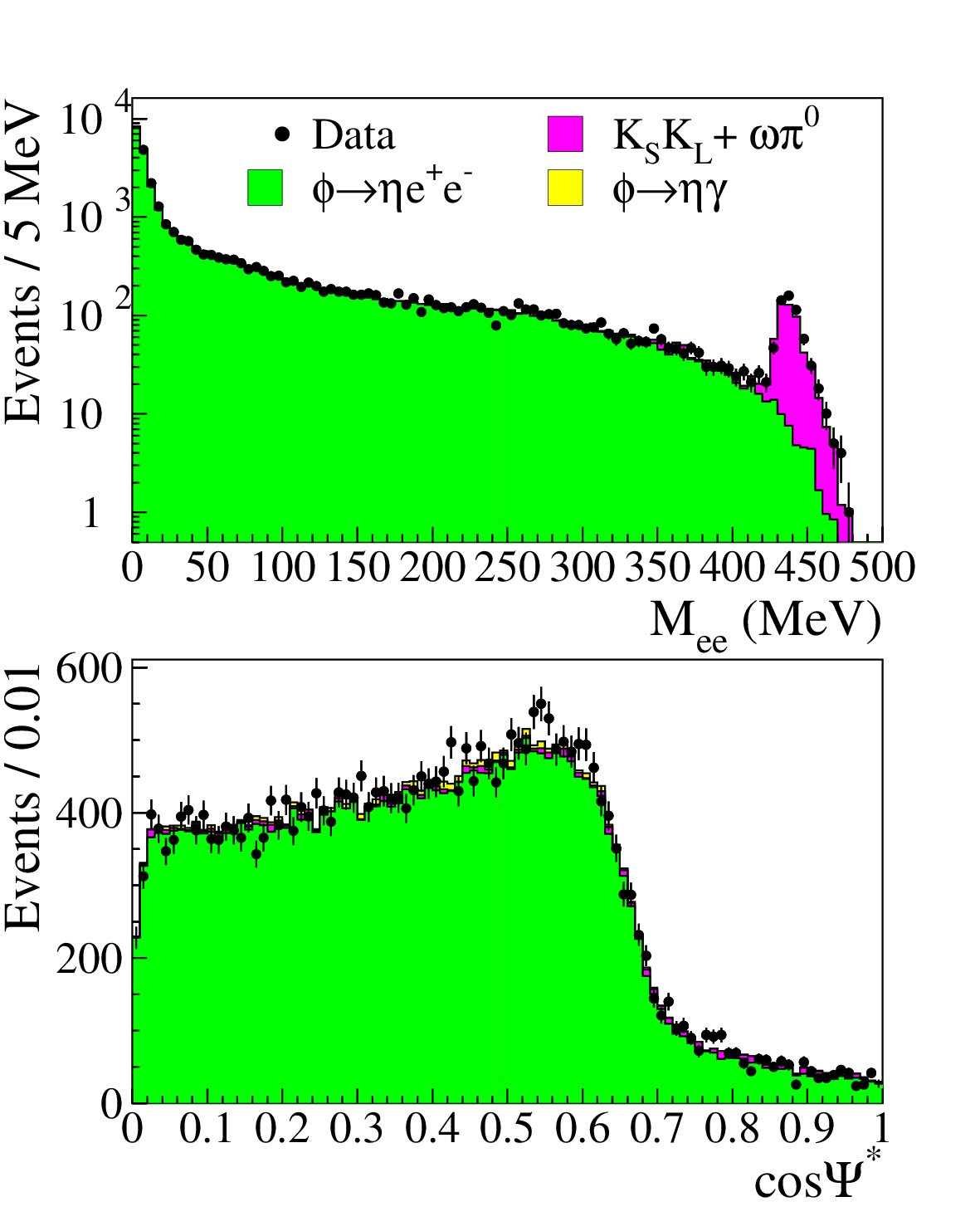,width=0.6\textwidth}
  \end{center}
  \caption{$\phi\to\eta\,\ep\el$, $\eta\to\pio\pio\pio$ events: 
    data-MC comparison for $M_{ee}$ (top) and $\cos\Psi^*$ 
    distributions (bottom) at the end of the analysis chain.}
  \label{Fig:CompNeuAfter}
\end{figure}

At the end of the analysis chain, 30577 events are selected, with 
$\sim 3\%$ background contamination (Fig.~\ref{Fig:CompNeuAfter}).
The analysis efficiency, defined as the ratio between events surviving 
analysis cuts and generated events, is $\sim 15\%$ at low \ep\el\ 
invariant masses, increasing up to $30\%$ at higher $M_{ee}$ values.

\begin{figure}[!t]
  \begin{center}
    \epsfig{file=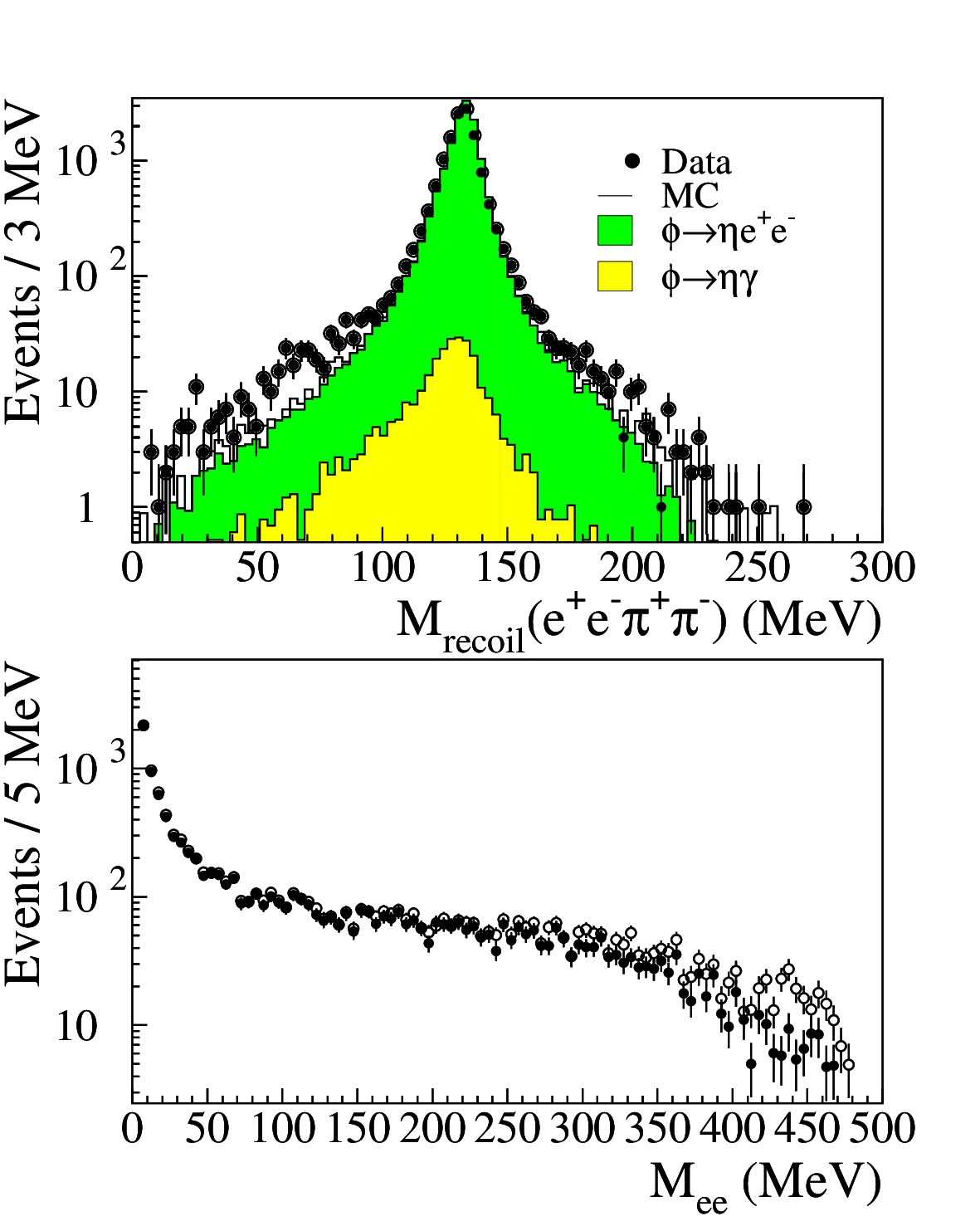,width=0.6\textwidth}
  \end{center}
  \caption{$\phi\to\eta\,\ep\el$, $\eta\to\pip\pim\pio$ analysis.
    Top: data-MC comparison for the recoil mass against the 
    $\ep\el\pip\pim$ system. Bottom: $M_{ee}$ distribution before 
    (open circles) and after (black dots) the cut on 
    $M_{\rm recoil}(ee\pi\pi)$.}
  \label{Fig:Charged}
\end{figure}

The analysis of the decay channel $\eta\to\pip\pim\pio$
is the same as described in \cite{UbosonKLOE}, with the addition of
a cut on the recoil mass to the $\ep\el\pip\pim$ system, which is 
expected to be equal to the \pio\ mass for signal events. 
In Fig.~\ref{Fig:Charged} top, data-MC comparison shows some residual 
background contamination in the tails of the distribution, which are
not well described by our simulation. 
A cut $100<M_{\rm recoil}(ee\pi\pi)<160$ MeV is then applied. The effect
of this cut on the $M_{ee}$ variable is shown in Fig.~\ref{Fig:Charged} 
bottom. The total number of selected events is 13254, with $\sim 2\%$ 
background contamination.

\section{\boldmath Upper limit evaluation on $U$ boson production}

\begin{figure}[!t]
  \begin{center}
    \epsfig{file=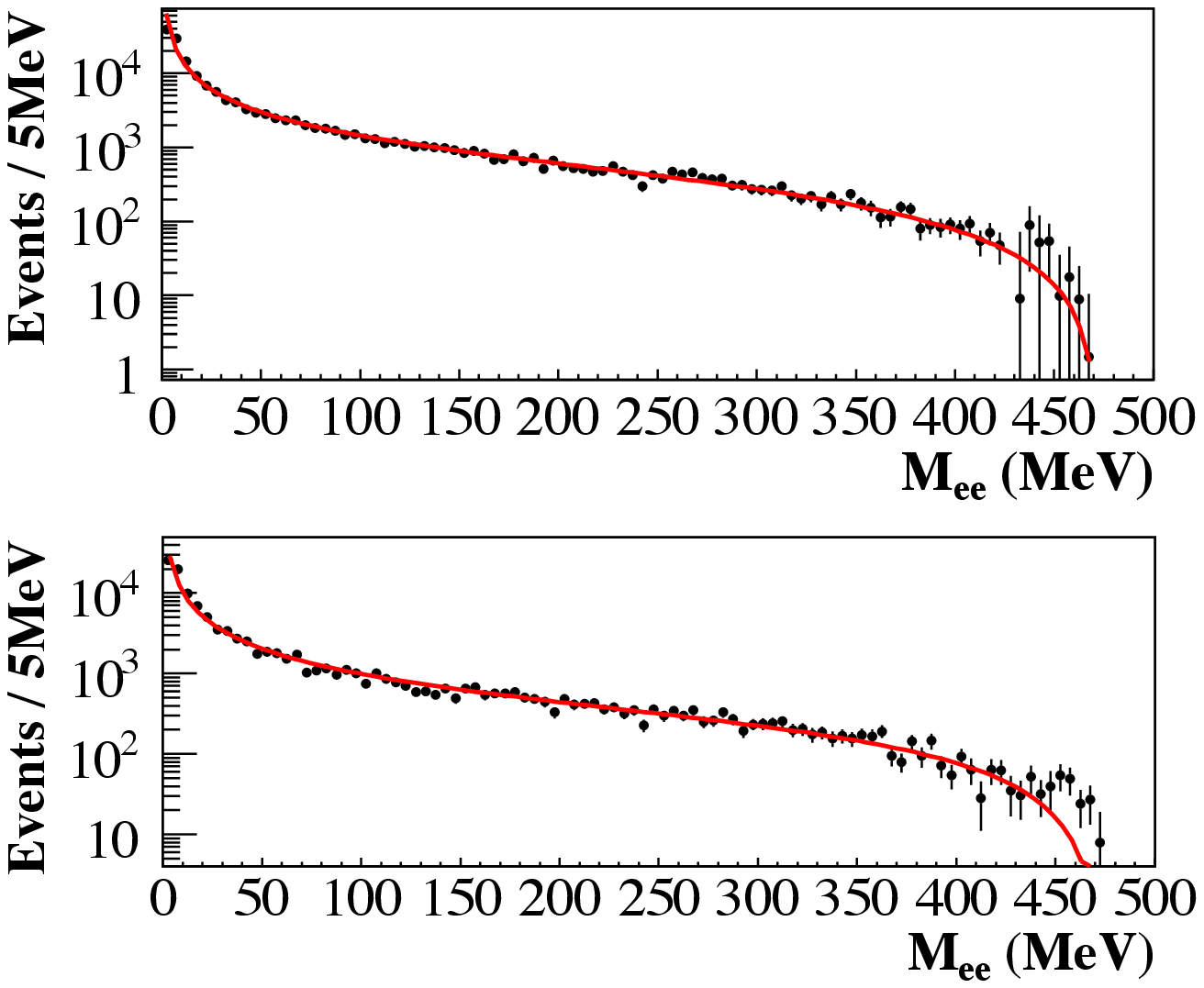,width=0.6\textwidth,angle=90}
  \end{center}
  \caption{Fit to the corrected $M_{ee}$ spectrum for the Dalitz 
    decays $\phi\to\eta\,e^+e^-$, with $\eta\to\pio\pio\pio$ (top) 
    and $\eta\to\pip\pim\pio$ (bottom).}
  \label{Fig:Fit}
\end{figure}

The upper limit on the $U$ boson production in the $\phi\to\eta U$
process is obtained combining the two $\eta$ decay channels. 
The resolution of the \ep\el\ invariant mass has been evaluated with
a Gaussian fit to the difference between the reconstructed and generated
mass for Monte Carlo events, providing $\sigma_{M_{ee}}\leq 2$ MeV over
the whole $M_{ee}$ range. 
The determination of the limit is done by varying the $M_U$ mass, 
with 1 MeV step, in the range between 5 and 470 MeV. Only 
five bins (5 MeV width) of the reconstructed $M_{ee}$ variable,
centered at $M_{U}$ are considered.
For each channel, the irreducible background, $b(M_{U})$, is extracted 
directly from our data after applying a bin-by-bin subtraction of the 
non-irreducible backgrounds and correcting for the analysis efficiency.
The $M_{ee}$ distribution is then fit, excluding the bins used for the 
upper limit evaluation. The parametrization of the fitting function has 
been taken from Ref.~\cite{Landsberg85}. The $\phi\eta\gamma^*$ 
transition form factor is parametrized as
\begin{equation}
  F_{\phi\eta}(q^2) = \frac{1}{1-q^2/\Lambda^2}
 \label{Eq:FF}
\end{equation}
with $q=M_{ee}$. Free parameters are $\Lambda$ and a normalization factor.
The spread of the extracted parameters is contained within the
statistical error of the fit done on the whole $M_{ee}$ mass
range, shown in Fig.~\ref{Fig:Fit}, as expected from the overall 
good description of the $M_{ee}$ shape for both $\eta$ decay channels.

\begin{figure}[!t]
  \begin{center}
    \epsfig{file=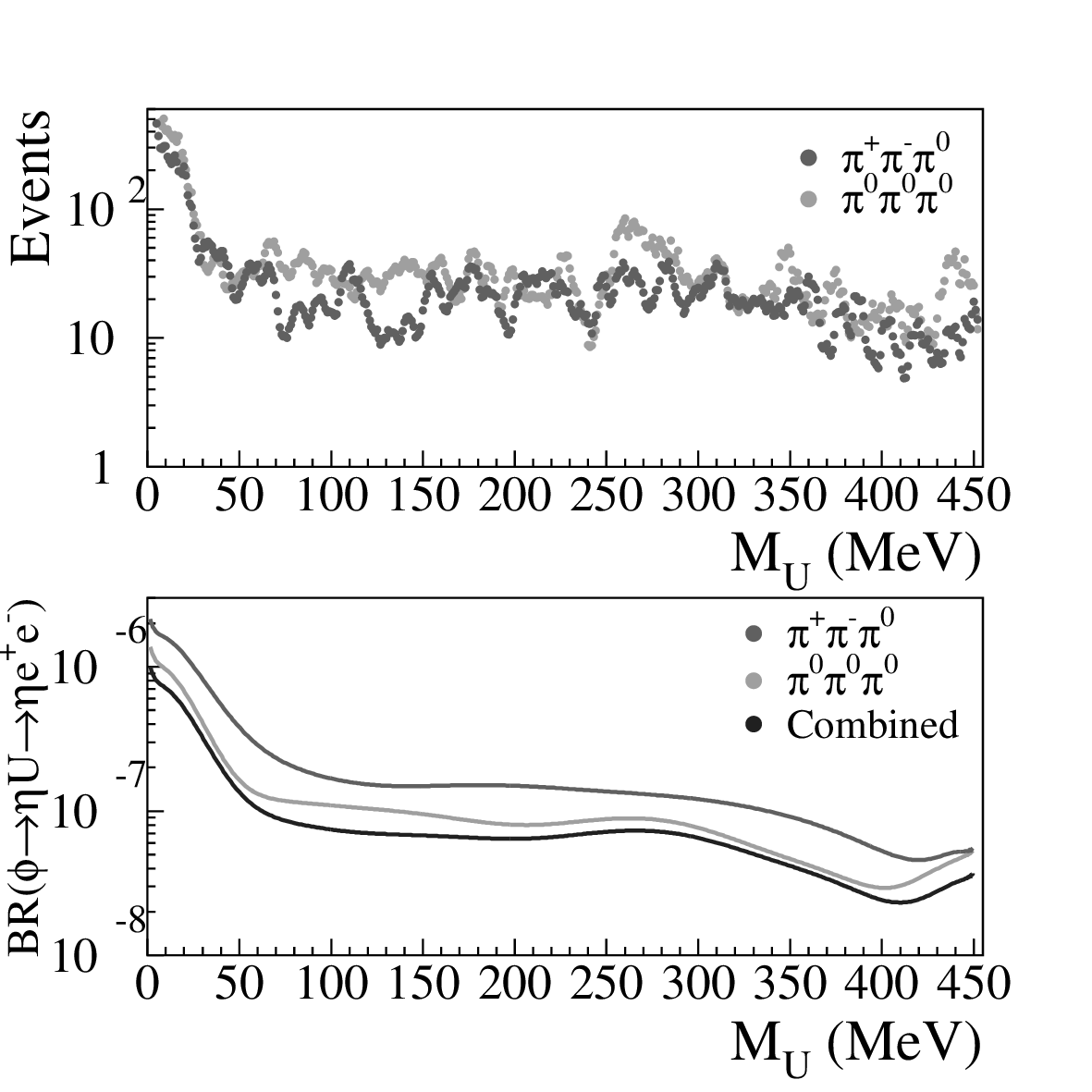,width=0.6\textwidth} 
  \end{center}
  \caption{Top: upper limit at 90\% C.L. on the number of events 
    for the decay chain $\phi\to\eta\,U$, $U\to\ep\el$, with 
    $\eta\to\pio\pio\pio$ and $\eta\to\pip\pim\pio$. Bottom: 
    smoothed upper limit at 90\% C.L. on 
    ${\rm BR}(\phi\to\eta\,U)\times{\rm BR}(U\to\ep\el$), 
    obtained separately for the two $\eta$ decay channels and 
    from the combined analysis.}
  \label{Fig:UL_nev}
\end{figure}

The exclusion limit on the number of events for the $\phi\to\eta\, U$ 
signal as a function of $M_{U}$ is obtained with the CL$_{\rm S}$ technique
\cite{CLS}, using the $M_{ee}$ spectra before background subtraction.
The limit is extracted both for each $\eta$ decay channel and in a 
combined way. For the combined procedure, the CL$_{\rm S}$ evaluation 
is done by summing values over all bins of the two decay channels, 
taking into account the different luminosity, efficiency and relative 
branching ratios of the two samples.
The systematic error on the background knowledge $\Delta b(M_{ee})$ 
is evaluated, for each $M_U$ value, changing by one standard deviation 
the two fit parameters and has been taken into account 
while evaluating CL$_{\rm S}$, applying a Gaussian spread of width 
$\Delta b(M_{ee})$ on the background distribution.
In Fig.~\ref{Fig:UL_nev} top, the upper limit at 90\% C.L.\ on 
the number of events for the decay chain $\phi\to\eta\,U$, $U\to\ep\el$, 
is shown for both $\eta\to\pio\pio\pio$ and $\eta\to\pip\pim\pio$, 
separately evaluated.
In Fig.~\ref{Fig:UL_nev} bottom, the smoothed upper limit on the 
branching fraction for the process $\phi\to\eta\,U$, $U\to\ep\el$, 
obtained from the combined method is compared with evaluations from
each of the two decay channels. 
In the combined result, the upper limit on the product 
${\rm BR}(\phi\to\eta\,U)\times{\rm BR}(U\to\ep\el$) varies from $10^{-6}$ 
at small M$_U$ to $\sim 3\times 10^{-8}$ at 450 MeV.

The exclusion plot in the $\alpha'/\alpha=\epsilon^2$ {\it vs} M$_U$ 
plane, where $\alpha'$ is the coupling of the $U$ boson to electrons 
and $\alpha$ is the fine structure constant, has been finally derived 
assuming the relation \cite{Reece:2009un}:
\begin{equation}
  \sigma(\ep\el\to\phi\to\eta\,U) = \epsilon^2 \, |F_{\phi\eta}(m_U^2)|^2 \,
  \frac{\lambda^{3/2}(m_\phi^2,m_\eta^2,m_U^2)}
       {\lambda^{3/2}(m_\phi^2,m_\eta^2,0)}
  \, \sigma(\ep\el\to\phi\to\eta\gamma) \, ,
  \label{Eq:Reece}
\end{equation}
with $\lambda(m_1^2,m_2^2,m_3^2) = [1 + m_3^2/(m_1^2-m_2^2)]^2 - 
4 m_1^2 m_3^2/(m_1^2-m_2^2)^2$.
We assume that the $U$ boson decays only to lepton pairs, with equal 
coupling to \ep\el\ and $\mu^+\mu^-$.

\begin{figure}[!t]
  \begin{center}
    \epsfig{file=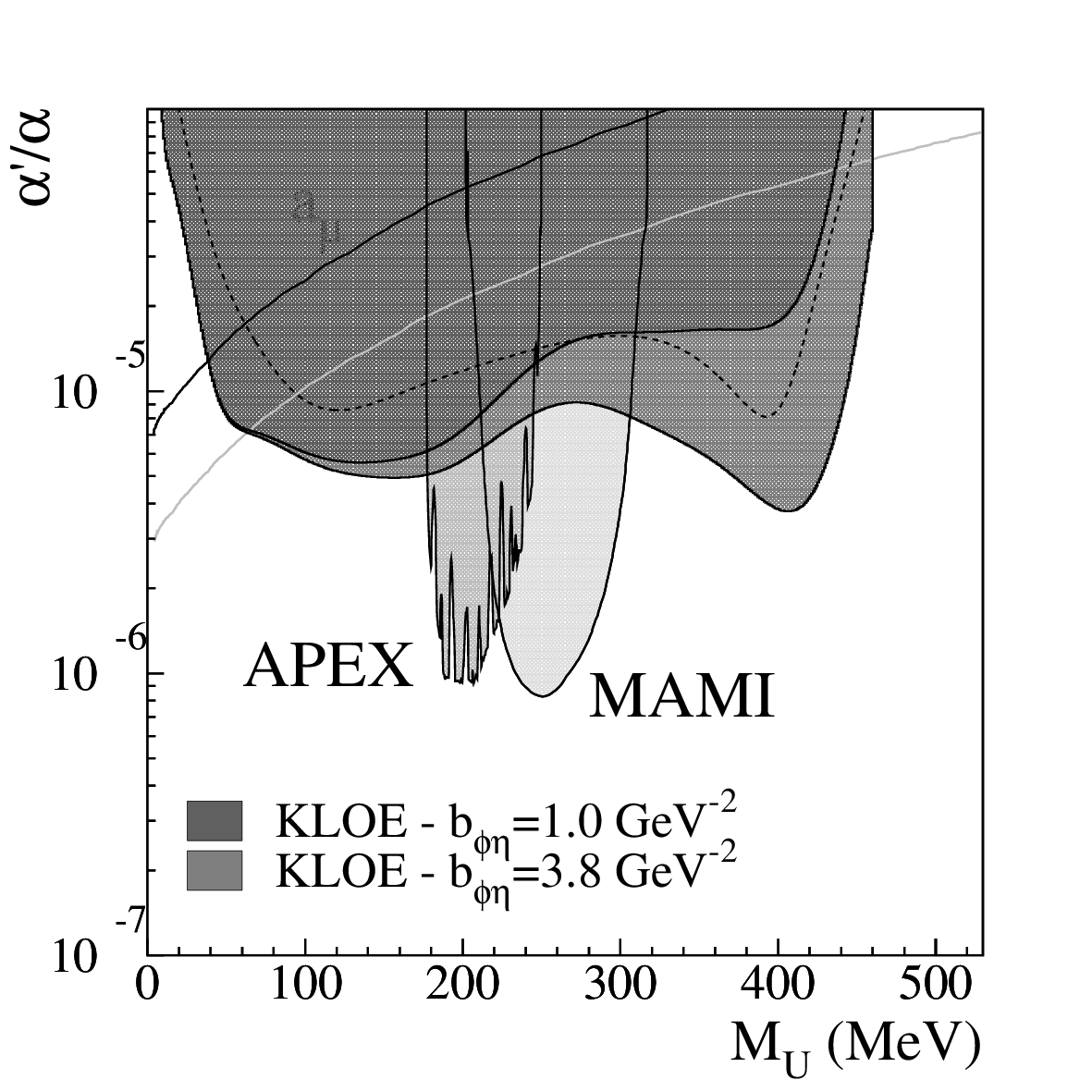,width=0.6\textwidth} 
  \end{center}
  \caption{Exclusion plot at 90\% C.L. for the parameter 
    $\alpha'/\alpha=\epsilon^2$, compared with existing limits from
    the muon anomalous magnetic moment and from MAMI/A1 and 
    APEX experiments. The gray line represents the expected values 
    of the $U$ boson parameters needed to explain the observed 
    discrepancy between measured and calculated $a_\mu$ values. 
    The dotted line is the previous KLOE result, obtained with the 
    $\eta\to\pip\pim\pio$ channel only.}
  \label{Fig:ULcurves}
\end{figure}

The extraction of the limit on the $\alpha'/\alpha$ parameter is 
related to the parametrization of the form factor (Eq.~\ref{Eq:Reece}), 
and thus to the $\Lambda$ parameter in Eq.~\ref{Eq:FF}. The SND 
experiment measured the form factor slope, 
$ b_{\phi\eta} = {dF/dq^2|}_{q^2=0} = \Lambda^{-2}$,
obtaining $b_{\phi\eta} = (3.8\pm 1.8)$ GeV$^{-2}$ \cite{phietaeeSND}, 
with a central 
value different from theoretical predictions based on VMD 
($b_{\phi\eta} \sim 1$ GeV$^{-2}$) \cite{AchasovFF}, although in 
agreement within the error.
In Fig.~\ref{Fig:ULcurves} the smoothed exclusion plot at 90\% C.L.\ 
on $\alpha'/\alpha$ is compared with existing limits in the same 
region of interest \cite{MAMI,APEX,amu}. The evaluation is done 
using both the experimental and the theoretical values of the form 
factor slope. The two resulting curves overlap at low $M_{ee}$ 
values, while the limit obtained using the SND measurement gives 
an increasingly larger exclusion region up to $\sim 400$ MeV, 
moving closer to the other curve at the end of the phase space.
Having the experimental 
value of $b_{\phi\eta}$ an uncertainty of
$\sim 50\%$, we conservatively use the curve obtained 
with theoretical predictions, resulting in a limit of:
$\alpha'/\alpha < 1.7 \times 10^{-5}$ for $30<M_{U}<400$ MeV, and
even better for the sub-region $50<M_{U}<210$ MeV:
$\alpha'/\alpha < 8.0 \times 10^{-6}$.
Comparing our result with the previous KLOE measurement, reported
as the dotted line in Fig.~\ref{Fig:ULcurves}, we improve the upper 
limit of about a factor of two
when using the same parametrization of the form factor.
This result reduces the region of the $U$ boson parameters that 
could explain the observed discrepancy between the measurement and 
Standard Model prediction of the muon anomalous magnetic moment, 
$a_\mu$, ruling out masses in the range 60--435 MeV.

\section*{Acknowledgments}

We warmly thank our former KLOE colleagues for the access to the 
data collected during the KLOE data taking campaign.
We thank the DA$\Phi$NE team for their efforts in maintaining low 
background running conditions and their collaboration during all 
data taking. We want to thank our technical staff: G.F.~Fortugno and 
F.~Sborzacchi for their dedication in ensuring efficient operation 
of the KLOE computing facilities; M.~Anelli for his continuous attention 
to the gas system and detector safety; A.~Balla, M.~Gatta, G.~Corradi 
and G.~Papalino for electronics maintenance; M.~Santoni, G.~Paoluzzi 
and R.~Rosellini for general detector support; C.~Piscitelli for his 
help during major maintenance periods. 
This work was supported in part by the EU Integrated Infrastructure 
Initiative Hadron Physics Project under contract number RII3-CT- 2004-506078; 
by the European Commission under the 7th Framework Programme through 
the `Research Infrastructures' action of the `Capacities' Programme, 
Call: FP7-INFRASTRUCTURES-2008-1, Grant Agreement No. 227431; by the 
Polish National Science Centre through the Grants No. 0469/B/H03/2009/37, 
0309/B/H03/2011/40, DEC-2011/03/N/ST2/02641,\\ 2011/01/D/ST2/00748 and 
by the Foundation for Polish Science through the MPD programme and the 
project HOMING PLUS BIS/2011-4/3.


\end{document}